\documentclass[%
 reprint,
nobibnotes,
 amsmath,amssymb,
aps,
twocolumn
]{revtex4-2}

\usepackage{graphicx}% Include figure files
\usepackage{dcolumn}% Align table columns on decimal point
\usepackage{bm}% bold math

\begin{document}
\title{Self-starting microring laser solitons from the periodic cubic complex Ginzburg-Landau equation}

\author{Martin Francki\'e}

\date{\today} % Leave empty to omit a date

\begin{abstract}
Dissipative Kerr solitons are optical pulses propagating in a nonlinear dielectric waveguide without dispersing. These attractive properties have spurred much research into integrated soliton generation in microring resonators at telecom wavelenghts. However, these solitons are generated via external pumping of a nonlinear medium, which limits system compactness and the wavelength coverage. Recently it was shown that dissipative Kerr solitons can be generated directly in the gain medium of mid-infrared semiconductor ring lasers, where the energy is supplied via electrical pumping. These technological advances enable a new route towards the monolithic generation of diverse light states in a wide frequency range, requiring insight into the conditions necessary for achieving control over the system. In this work, we study in depth the phase diagram of such systems, exhibiting fast gain dynamics and no external pumping, and which are described by the complex cubic Ginzburg-Landau equation with periodic boundary conditions. While previous studies have focused on large system sizes, we here focus on finite ring sizes leading to a modification of the phase space and enlargement of the region for stable soliton formation. In addition to localized, dispersion-less soliton pulses on a finite background, breather solitons are predicted to occur under feasible experimental conditions. This enables monolithic generation of solitions in electrically pumped media such as quantum cascade lasers, extending the availability of soliton sources to mid-infrared and terahertz frequencies which are very attractive for molecular spectroscopy, free-space communication, imaging, and coating thickness measurements.
\end{abstract}

\maketitle

\section{Introduction}%
Dissipative Kerr solitons (DKS) are electro-magnetic pulses propagating through a dielectric waveguide without dispersing.\cite{grelu_dissipative_2012} They can be generated via four-wave mixing in optically pumped passive ring cavities with a large Kerr nonlinearity and anomalous dispersion\cite{kippenberg_dissipative_2018}, where the latter compensates the intensity-dependent frequency shifts induced by self- and cross-phase modulation. The balancing of dispersion and non-linear frequency shifts in these periodic systems leads to equidistant modes in the frequency domain, and thus microding DKS are frequency combs with a linear phase relation between neighboring modes. They have recently been realized in a variety of materials with large nonlinear coefficients, such as CMOS-compatible silicon, silicon nitride, and silicon dioxide\cite{kippenberg_dissipative_2018}, as well as the III-V semiconductor AlGaAs\cite{moille_dissipative_2020}, favourable for its large nonlinear coefficient\cite{chang_ultra-efficient_2020}. Soliton frequency combs have recently found practical applications in optical communication\cite{marin-palomo_microresonator-based_2017}, and are promising candidates for squeezed light generation and quantum information processing.\cite{arrazola_quantum_2021} At the same time, different waveguide configurations are actively researched for improving soliton characteristics.\cite{mittal_topological_2021}
%extract background-free solitons, increase the conversion efficiency, exploit the robustness of topological photonics\cite{mittal_topological_2021}, etc.

 Despite their success, the generation of optically pumped microcavity solitons also come with certain challenges\cite{kippenberg_dissipative_2018}. For instance, they require cavities with extremely high quality factors (Q-factors) in order for the parametric gain to overcome the linear loss, which, together with the modest nonlinearity in the conventionally used material systems, severely limits their overall efficiency to at most a few percent\cite{bao_nonlinear_2014,xue_microresonator_2017}. Notably, recent suggestions with more elaborate waveguides with photonic molecules claim to have achieved significantly higher efficiency.\cite{helgason_power-efficient_2022} Furthermore, the external pump is imprinted also at the output, which requires very narrow spectral filtering which may be undesirable for many applications, and complicating photonic integration. Finally, solitons in the mid- and far-infrared regions have not yet been realized in such systems, which have been mostly realized at telecom frequencies. Integrating an optical gain medium directly inside of a ring laser cavity could provide solutions to these challenges, as already indicated by the first mid-infrared soliton being achieved in such a system\cite{meng_dissipative_2021}.

\begin{figure}
    \centering
    \includegraphics[width=\linewidth]{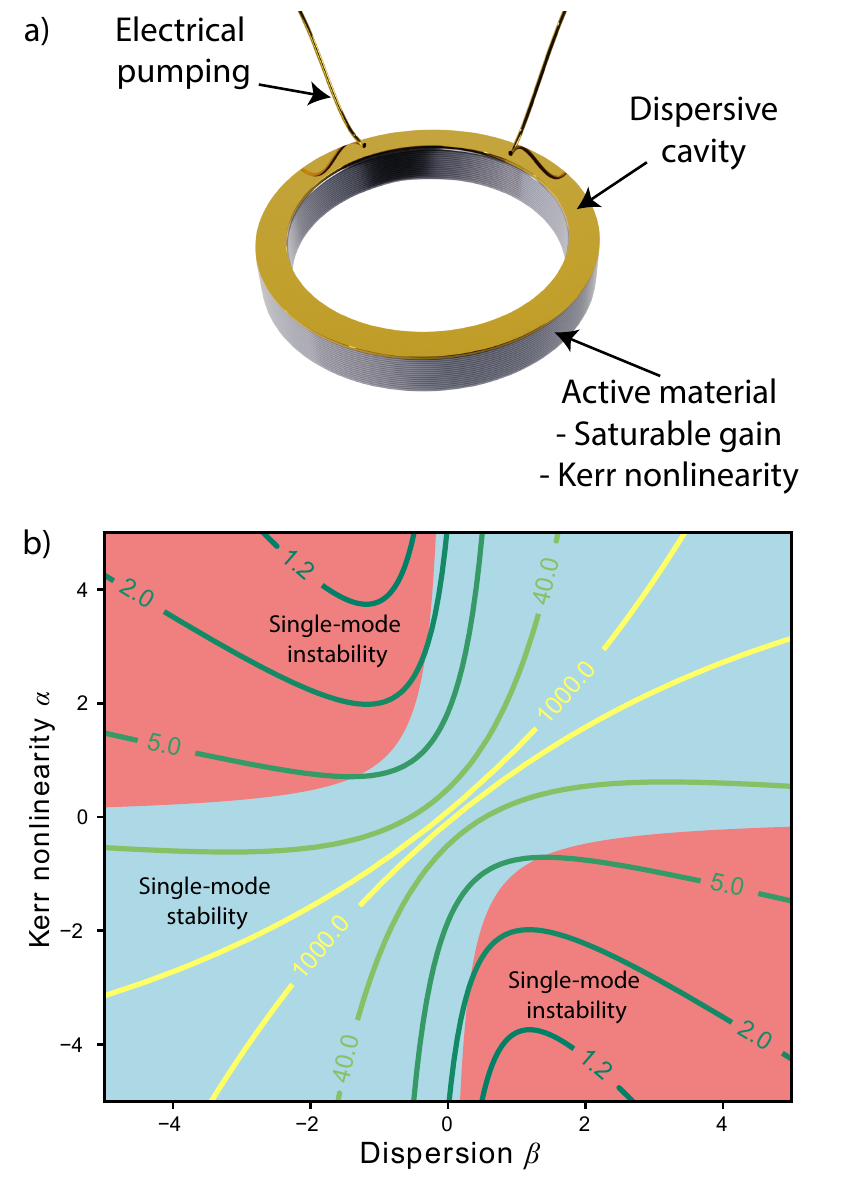}
    \caption{Stability regions for the ring laser schematically shown in (a), with stable single-mode solutions shaded in blue and multi-mode instability regions shaded in red. The contour lines show the (normalized) widths of the analytical soliton solutions to the CGLE. The data shown in b) are derived from an infinitely extended ring cavity, with $\alpha$ and $\beta$ defined in Eq.~\eqref{eq:norm_CGLE}.}
    \label{fig:bistability_2d}
\end{figure}

The dynamics of DKS in optically pumped Kerr microrings cavities are well understood from the Lugiato-Lefever equation (LLE)\cite{chembo_modal_2010, chembo_spatiotemporal_2013, herr_temporal_2014}. Very similar behaviour as in dissipative Kerr microcavities was simulated when a net gain was present inside the optically pumped cavity\cite{columbo_unifying_2021}, in spite of the detrimental effect of a fast saturable gain on the generation of pulses.
%Such an optically pumped active ring laser\cite{gustave_dissipative_2015} was studied using a Maxwell-Bloch formalism and a generalized LLE in Ref.~\cite{columbo_unifying_2021}.
%However, these approaches do not include the full frequency-dependence, and thus the finite response time, of the light-matter coupling. Thus, they cannot be used for quantitatively studying real devices.
 %There, soliton formation was simulated for a narrow range of parameters with an external continuous-wave pump, which results in a similar bistability diagram as in the dissipative case, as is indicated by the yellow line in Fig.~\ref{fig:bistability_2d} b). By varying the external pump intensity, a similar transition from single-mode, to Turing rolls and unstable patterns, to cavity solitons, as in dissipative Kerr microcavities was simulated, in spite of the detrimental effect of a fast saturable gain.
In contrast, in an active ring laser without optical pumping, the electric field is governed by the cubic complex Ginzburg-Landau equation (CGLE)
\begin{equation}
    \dot{E} = (g + \text{i}\theta)E + (\Delta + \text{i}\gamma) |E|^2E + (\epsilon + \text{i}\delta)\nabla^2 E, \label{eq:general_CGLE}
\end{equation}
where the real and imaginary parts of the complex coefficients describe gain ($g$) and detuning ($\theta$), gain saturation ($\Delta$) and Kerr nonlinearity ($\gamma$), as well as gain curvature ($\epsilon$) and dispersion ($\delta$). This equation is derived from the Maxwell's equations for a fast saturable gain medium in Appendix \ref{app:CGLE_derivation}. This is a widely studied nonlinear differential equation in physics, describing a plethora of phenomena such as superconductivity and second-order phase transitions\cite{aranson_world_2002}, and ocean waves\cite{benjamin_disintegration_1967}. Previous studies\cite{shraiman_spatiotemporal_1992, bretherton_intermittency_1983, chate_spatiotemporal_1994, akhmediev_singularities_1996} have mainly focused on extended (infinite) systems where some analytical results can be obtained. Instead, this work focuses on periodic systems of finite size, such as microring lasers. Any periodic solutions, such as solitions, in this system have to balance gain and nonlinear gain saturation, in addition to dispersion and Kerr nonlinearity.

Written on it's normalized form:
\begin{eqnarray}
\frac{\partial \mathcal{E}}{\partial \tau} =
\mathcal{E} - (1 + \text{i}\alpha )|\mathcal{E}|^2\mathcal{E} + (1 + \text{i}\beta)\frac{\partial^2 \mathcal{E}}{\partial \Theta^2},
\label{eq:norm_CGLE}
\end{eqnarray}
the CGLE in the limit of a large cavity has two main parameter regions with distinct solutions, separated by the Benjamin-Feir (BF) instability line\cite{benjamin_disintegration_1967} $|\alpha\beta| = 1$ and shown in Fig.~\ref{fig:bistability_2d}b): for $|\alpha\beta| < 1$ these are plane-wave solutions\cite{benjamin_disintegration_1967}, while the region $|\alpha\beta| > 1$ is dominated by chaos, with weakly amplitude-modulated solutions in the phase-turbulence regime, and strongly amplitude-modulated solutions in the amplitude-turbulent regime\cite{shraiman_spatiotemporal_1992, chate_spatiotemporal_1994}. In addition to the single-mode solutions, analytical soliton solutions also exist\cite{akhmediev_singularities_1996, pereira_nonlinear_1977}, although these are not stable in the presence of gain as discussed below. In addition, the larger the $|\alpha\beta|$, i.~e.~the further from the BF instability the laser is operated, the fewer attractors exist\cite{shraiman_spatiotemporal_1992}.  While the solutions mentioned above were studied for spatially extended systems, one would expect soliton solutions in the region where $|\alpha\beta| \gtrsim 1$ also for the sysem with finite periodicity, although the phase-diagram structure may be qualitatively different and depend on the spatial extent of the system. Indeed, for a ring QCL simulated using the Maxwell-Bloch formalism\cite{piccardo_frequency_2020}, weakly amplitude-modulated, periodic phase-turbulent solutions\cite{chate_spatiotemporal_1994} were found in this region with a spectrum reminiscent of the sech$^2$ fall-off characteristic of the analytical soliton solution, and was also found experimentally\cite{piccardo_frequency_2020}. However, further investigation into the BF instability region was not performed, and no soliton solutions with significant amplitude modulation were found.

In contrast, very recently solitons were observed in a ring QCL\cite{meng_dissipative_2021}, showing conclusively that the formation of strongly amplitude-modulated frequency combs with short pulses are possible in this geometry. Using a mode-expansion model based on susceptibilities, including the frequency-dependence of the gain, a similar behaviour could be simulated\cite{meng_dissipative_2021}. It is thus of great interest and importance to further the investigation into the conditions for soliton formation in ring cavity lasers. To this end, an extensive exploration of the parameter space is key. Up until now, however, no systematic study of the solutions to the periodic CGLE beyond the BF instability has been conducted. In the present work, we propose a formalism based on susceptibilities and Maxwell's equations, described by the CGLE, for modeling the electric field inside ring cavity lasers. (The same form of the equation can also be derived from the Maxwell-Bloch equations for a two-level system\cite{piccardo_frequency_2020} under certain approximations.) It predicts reliable generation of solitons of large amplitude modulation at the interface of the predicted single- and multi-mode regimes, in accordance with recent experimental and theoretical results\cite{meng_mid-infrared_2020,piccardo_frequency_2020,piccardo_frequency_2020, meng_dissipative_2021}. In addition, we discuss dynamical, periodic, solutions, akin to breather solitons in passive microcavities, which were previously not reported in this system. We also compare the numerical results to analytical soliton solutions to the CGLE with gain. Finally, we discuss the dependence of soliton formation on the gain curvature and ring size.

\section{Theory}

In order to study solutions in different parameter regimes, we fix the periodicity and vary the two parameters in the normalized CGLE \eqref{eq:norm_CGLE}.
%derive the general form of the CGLE \eqref{eq:general_CGLE} for the electric field amplitude $E$ in Appendix \ref{app:CGLE_derivation}.
%To provide a context for Eq.~\eqref{eq:general_CGLE}, the physical difference of this equation to the (dissipative) LLE case is highlighted in Fig.~\ref{fig:bistability_2d}, where we show the stability of homogeneous solutions to Eq.~\eqref{eq:general_CGLE} with an additional driving term (as in Eq.~\eqref{eq:app_CGLE}). As is well known, the LLE has a bistability for sufficient red-detuning ($\theta > \sqrt{3}$ in Fig.~\ref{fig:bistability_2d} a)). On the other hand, the amplifying case has regions of two unstable and one stable branch, in addition to the bistable region marked by yellow stars in Fig.~\ref{fig:bistability_2d} b). Similar to the dissipative LLE, however, in the bistable region soliton-like solutions were found numerically in Ref.~\cite{columbo_unifying_2021}, and the behaviour was found to be similar to that of the purely dissipative cavity.
%We focus on the case where the energy provided to balance the optical loss is given directly from the gain medium, without any external pump field. In Fig.~\ref{fig:bistability_2d} b) these are the unstable zero-field solutions along the y-axis, as well as the stable finite-field solution marked by the red arrow in Fig.~\ref{fig:bistability_2d} b).
Solitons cannot be obtained in the regular procedure employed for Kerr microring resonators, namely detuning the external pump laser until the intracavity field leaves the stable, homogenous solution, which promotes inhomegenous solutions. Only by changing the internal parameters of the laser must it reach the unstable BF regime. This happens when the side modes experience enough parametric gain to overcome the cavity losses, which happens as the power of the single mode reaches above the parametric threshold\cite{meng_dissipative_2021, herr_temporal_2014}, and solitons emerge spontaneously. %\textbf{Note that this is different from the Risken-Nummedal-Graham-Haken instability threshold in Fabry-Perot lasers.}

Below the parametric threshold, the electric field corresponds to single-mode lasing, where gain saturation exactly balances the losses, and the self-phase modulation provides a frequency shift from the cold cavity modes by $\theta = -\alpha g$, where $\alpha$ is the negative of the linewidth enhancement factor (LEF). Increasing the Kerr nonlinearity (and the gain saturation parameter in order to keep the LEF constant) eventually results in a multi-mode instability, as the parametric gain on the side-modes outweigh the combined linear and parametric losses induced by the gain saturation. In a semiconductor laser, this happens as the bias is increased, leading to an increase in the inversion to which $\gamma$ and $\Delta$ are proportional.
\iffalse
Expanding the field in frequency modes we obtain the coupled-mode equations
\begin{eqnarray}
\dot{E}_m &=& \left(\text{Re}\{B_m\}) - k_m^2\text{Re}\{D_m\}\right)E_m \\&+& \text{i}\left( \text{Im}\{B_m\}E_m - k_m^2\text{Im}\{D_m\}\right) E_m  \nonumber \\
&+& \sum_{kl} C_{klm} E_k^\ast E_l E_{m + k - l},
\end{eqnarray}
where we have given mode indices to all constants of Eq.~\eqref{eq:general_CGLE} to allow for explicit frequency dependence. Here, it is clear that the spectral filtering $\epsilon $ provides a quadratic gain profile, which can be replaced by a more accurate Lorentzian profile of the frequency-dependent gain coefficients $\text{Re}\{B_m\}$, and quadratic dispersion is given by $\delta $. Compared to a spatio-temporal model, the mode expansion is efficient when the number of modes is small, but become computationally heavy for many $N\gtrsim 100$ modes due to the nonlinear term. The expected scaling is $N^3$ for a naive implementation, however the result will vary greatly due to the adaptive step-size used in the employed fourth-order Runge-Kutta solver. Empirically, we find our program scales as $\sim N^{2.5}$.
\fi

%Since we are interested in the situation where the laser is driven only by the internal gain, in the following we ignore the external drive $F$, corresponding the stable point (2,2) in Fig.~\ref{fig:bistability_2d} b). Returning to the time domain,
\subsection{Normalized CGLE}%
Eq.~\eqref{eq:general_CGLE} can be re-written in the normalized form of Eq.~\eqref{eq:norm_CGLE}
by transforming to a frame rotating with angular frequency $\theta$ and re-scaling the electric field, time, and coordinate according to $\mathcal{E} = E\sqrt{-\Delta/g }e^{\text{i}\theta  t}$, $\tau = t g $, $\Theta = z\sqrt{g /\epsilon }$,
$\beta = \delta /\epsilon $,
$\alpha = \gamma/\Delta$,
$\alpha$ being the negative of the LEF as before. The rotating frame transformation implicitly assumes a uni-directional propagating wave, and so the laser is henceforth assumed to operate above the symmetry-breaking point\cite{meng_dissipative_2021}. Eq.~\eqref{eq:norm_CGLE} only has two parameters $\alpha$ and $\beta$, which control respectively the ratio of Kerr nonlinearity to gain saturation and dispersion to spectral filtering. In a ring cavity, however, Eq.~\eqref{eq:norm_CGLE} has in addition the periodic boundary condition $\mathcal{E}(\Theta + \Pi) = \mathcal{E}$, where $\Pi = L \sqrt{g /\epsilon }$ and $L$ is the circumference of the ring. Again, the frequecy dependence of $\alpha$ and $\beta$ is neglected, and the sign of $\alpha$ will depend on the detuning from the gain transition and can be designed to some extent in a heterostructure gain medium (in contrast to the passive and dissipative case where the transition is far from the plasma frequency). $\beta > 0$ for anomalous dispersion, while the spectral filtering effectively limits the gain bandwidth to $\Pi/\pi$ modes with non-negative gain, since the effective gain on mode $m$ is
\begin{equation}
    g_m = g  -
    \epsilon k_m^2
    =
    g  - \epsilon  \frac{m^2}{R^2},
    \label{eq:spect_filt}
\end{equation}
where $k_m = 2\pi m/L$ is the mode's wave vector. Thus, we can explore all possible parameters by varying $\alpha$ and $\beta$, for a chosen number of modes inside the gain bandwidth. This can correspond to the distance from threshold, or the size of the ring, depending on how the non-normalized parameters are chosen.

%For the experiment in Ref.~\cite{Meng_NatPhot} with integrated second-order dispersion of $50$ kHz and a gain curvature of $\sim 5.5$ m${}^2$/s, the normalized parameters are $\alpha \approx 3.2$ and $\beta = 0.4$.

\subsection{Analytical solutions}
As we show in the Supplementary materials, Eq.~\eqref{eq:norm_CGLE} has analytical solutions
\begin{equation}
    \mathcal{E}(\Theta, t) = L \text{sech}(K\Theta)^{1 + \text{i}a}e^{-\text{i}\Omega t},
    \label{eq:sech_ansatz}
\end{equation}
with width $K^{-1} = \sqrt{a^2 + 2\beta a -1}$, frequency shift $\Omega = K^2(a^2\beta - \beta - 2a)$, magnitude $L = K\sqrt{a^2 + 3\beta a -2}$, and where
\begin{eqnarray}
a &=& - B \pm \sqrt{B^2 + 2} \label{eq:aeq} \\
B &=& \frac{3}{2}\frac{\alpha\beta + 1}{\alpha - \beta}.
\end{eqnarray}
Note, however, that this solution is strictly only valid for non-periodic boundary conditions and could be unstable as the tails go asymptotically to zero intensity, which is in general an unstable solution to Eq.~\eqref{eq:norm_CGLE}. Empirically, both through experimentation and modelling\cite{meng_dissipative_2021}, we have found that the solution to Eq.~\eqref{eq:norm_CGLE} rather resembles a pulse on a constant background or a sum of two out-of-phase pulses. However, such Ansätze result in transcendental equations, rather than closed-form analytical solutions, and the search for stable analytical solutions remains a difficult challenge. Nevertheless, as shown in Fig.~\ref{fig:analytical_width} (b), Eq.~\eqref{eq:sech_ansatz} can be used to estimate width and peak intensity of amplitude modulated solutions. The region of existence is given by the conditions $K^2 > 0$ and $|L|^2 > 0$, which can be fulfilled for all values of $\alpha$ and $\beta$ by choosing the appropriate sign in Eq.~\eqref{eq:aeq}. In Fig.~\ref{fig:analytical_width} a) the contour lines show the analytical soliton width $K^{-1}$ as a function of $\alpha$ and $\beta$, showing the balance between Kerr effect and dispersion as opposite sign between $\alpha$ and $\beta$ are required.

\section{Results}

\begin{figure*}
    \centering
    \includegraphics[width=\linewidth]{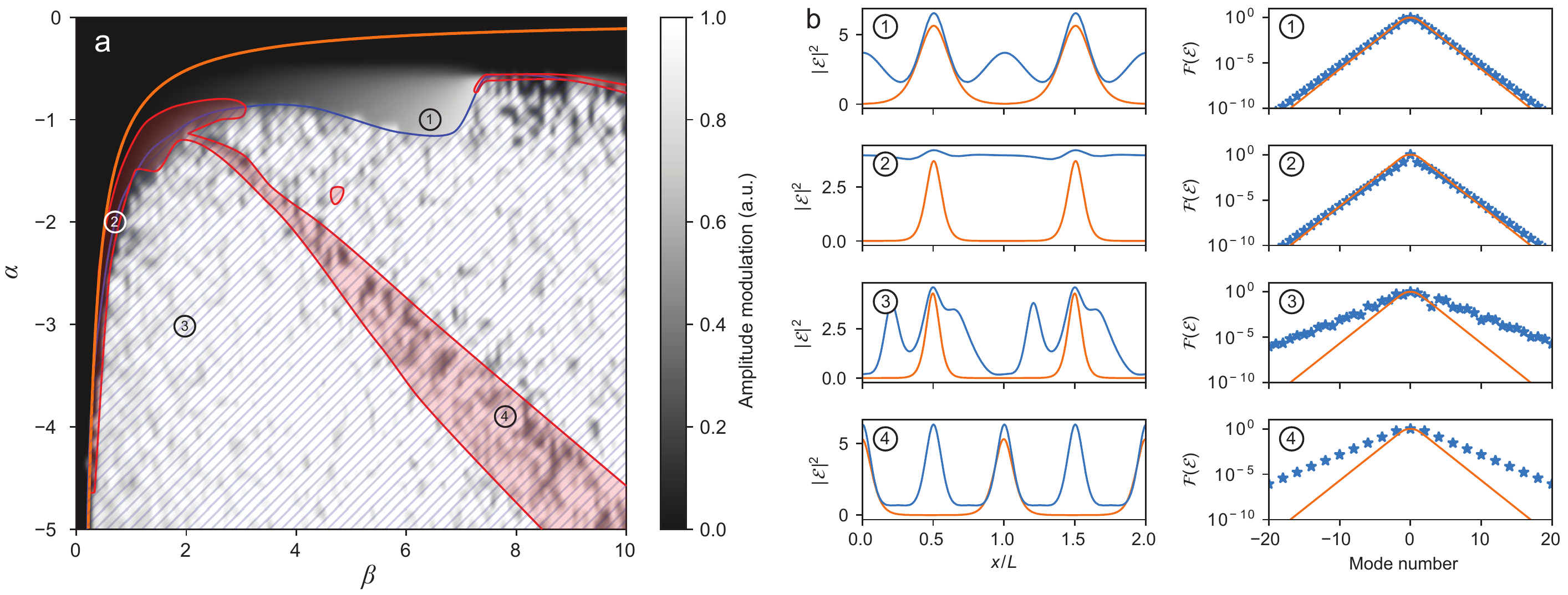}
    \caption{a) Phase diagram of numerical solutions (gray scale: single mode to soliton, red shaded regions: breather solitons, blue shaded region: chaos) as functions of the normalized parameters $\alpha$ and $\beta$ for $\Pi = 20$. The orange line shows the Benjamin-Fier instability $|\alpha\beta| = 1$ from Fig.~\ref{fig:bistability_2d}c). The simulations were performed for a duration of $\tau = 1000$, starting with the analytical soliton solution as initial condition. b) Intensity profiles and spectra after $\tau = 800$ of the numerical (orange) and analytic (blue) solutions to Eq.~\eqref{eq:norm_CGLE}. }
    \label{fig:analytical_width}
\end{figure*}

We have solved Eq.~\eqref{eq:norm_CGLE} for various ring sizes, defined by the number of modes with positive unsaturated gain. The qualitative picture shown in Fig.~\ref{fig:analytical_width} b) for $\Pi = 20$, and exhibits single-mode operation, frequency combs, breather solitons, and chaotic solutions. Clearly distinguishable regions appear for these different classes, where the largest areas contain chaotic or single-mode solutions, roughly separated by the Benjamin-Feir (BF) line. However, one can observe significant differences with regards to the case of extended ring cavities. Frequency combs appear in the pocket region with $\alpha \sim -1$ and $\beta$ in the range of approximately 6-7, as well as close to the BF line for $\beta < 1$ and $\alpha < -1$. A selection of solutions are compared to the analytic solutions for the same parameters in Fig.~\ref{fig:analytical_width} b). Compared to the analytical solution, the soliton solution labeled (1) has similar width and amplitude, but never reaches close to zero intensity between the pulse peaks. Instead, these soliton solutions appear on a background which grows as the pulse width narrows, thus stabilizing the solution in the tails of the pulse. The amount of amplitude modulation varies and increases with $|\alpha|$ and $|\beta|$, until the dispersion or Kerr nonlinearity becomes too large and the system undergoes an abrupt phase transition into the chaotic regime. As shown in Fig.~\ref{fig:analytical_width} b), the spectra of the soliton modes follow closely the $\text{sech}$ envelop of the corresponding analytical solution in regions (1) and (4). In contrast, for low $\beta$ and large $|\alpha|$, the narrow width of the analytical solution cannot be stabilized. For the parameters (2), corresponding to the mid-infrared ring QCL in Ref.~\cite{piccardo_frequency_2020}, the model replicates the frequency-comb solution with small amplitude modulation found by the Maxwell-Bloch formalism. Finally, for the parameters (3) the electric field is chaotic in the amplitude turbulance regime, where the intensity can go to zero at points along the ring.

%On the other hand, for the parameters (3), which numerically resulted in soliton solutions for the mid-infrared soliton QCL in Ref.~\cite{meng_dissipative_2021}, the relative amplitude modulation depth is similar to what was found experimentally (see Supplementary materials of Ref.~\cite{meng_dissipative_2021}). In contrast to the region around the point (1) in Fig.~\ref{fig:analytical_width}a), these amplitude-modulated frequency comb solutions exist only for a narrow range of $\beta$ values.

%\newpage
%\onecolumngrid

%\twocolumngrid

\begin{figure}[h]
    \centering
    \includegraphics[width=\linewidth]{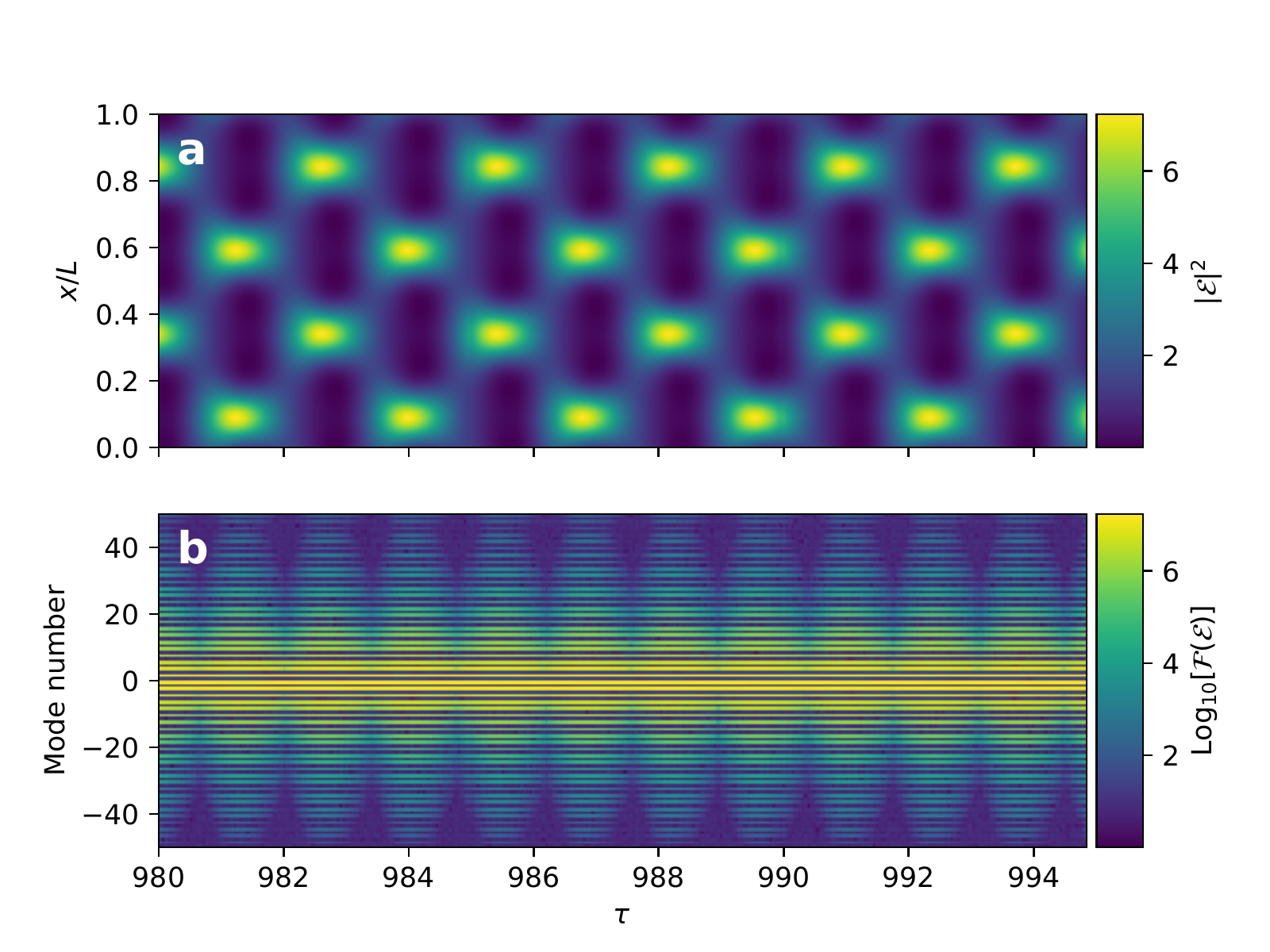}
    \caption{An example of a breather solution with parameters $\alpha = -2.5 $ and $\beta = 5.2$, corresponding to point (4) in Fig.~\ref{fig:analytical_width} a). The results are shown in the frame rotating with the soliton center frequency $\theta $ and shows only the relative motion of the pulses.}
    \label{fig:breather}
\end{figure}

Along the diagonal of Fig.~\ref{fig:analytical_width} a), dynamically stable solutions, analogous to breather modes in passive microcavities\cite{matsko_excitation_2012}, are consistently found. These solutions are characterized by a spectral width oscillating in time with a slow period compared to the round-trip time and the optical frequency. An example of this is shown in Fig.~\ref{fig:breather}, exhibiting two pairs of pulses propagating in the cavity with different group velocities. Every pulsation period, the pulses meet at two points on opposite sides of the ring cavity, resulting in an increased peak intensity which reduces the relative velocity of the pulses owning to the stronger nonlinear interaction\cite{ku_slow_2004}. This gives rise to a breathing standing-wave pattern of intra-cavity intensity, and would appear as a slowly modulated pulse train when observing the output of such a laser. The instantaneous spectrum when the peaks collide, shown in the lower right plot in Fig.~\ref{fig:analytical_width} b), shows only every second mode is occupied. This resembles harmonic frequency combs in Fabry-Perot lasers\cite{mansuripur_single-mode_2016}, where the harmonic state is attributed to spatial hole burning\cite{piccardo_harmonic_2018}, interaction between asymmetric optical transitions\cite{forrer_self-starting_2021} or optical feedback. In contrast, the wide frequency spacing seen for this state, which is not a frequency comb, results from the interference between the two opposing pulses in the cavity.

Finally, the behavior of the electric field is studied as the number of modes $N$ with net unsaturated gain is increased (corresponding to increasing ring radius or increasing the gain from the threshold). As $N$ is increased, the region around point (1) of stable solitons moves towards higher $\beta$ in Fig.~\ref{fig:analytical_width} a), while the unstable region approaches the BF instability line. For $N = 10$, the stable soliton region is outside of the studied range $\beta < 10$, but solutions with weaker amplitude modulation are still found close to the BF line.

\section{Conclusion}

In conclusion, self-starting intra-cavity gain solitions were demonstrated in ring cavities, using a spatio-temporal model of the CGLE. The model includes instantaneous gain saturation and a finite response time of the gain medium, in addition to dispersion and Kerr nonlinearity. Thus, recent experimental and numerical observations that solitions can form in purely electrically driven gain media have been confirmed with minimal model complexity. These results will enable further exploration of the phase space in order to guide future generations of compact, integrated pulsed and coherent sources. Furthermore, the presence of gain inside the cavity relaxes the stringent requirement of very high Q-factors and allows for immediate monolithic realization of solition sources with extended wavelength coverage down to the THz range. While background-free pulses are the analytic solutions to the modeled equation, they are inherently unstable due to the gain, and stable solutions appear to require a significant background. However, it should also be kept in mind that this system benefits from not having a constant background coming from a single-mode pump laser employed in passive dissipative Kerr micro-resonators.

Suitable gain media can be realized with many semiconductor laser systems, such as diode lasers\cite{sterczewski_frequency-modulated_2020}, quantum well lasers\cite{sorel_operating_2003}, quantum dot lasers\cite{lu_ultra-narrow_2018}, quantum cascade lasers (QCLs)\cite{faist_quantum_1994}, and interband cascade lasers (ICLs)\cite{meyer_type-ii_1996}, although for media with slow gain saturation the model would have to be supplemented with further memory terms, such as frequency-dependent and higher-order nonlinearities. Quantum wells and superlattices can be designed with huge dipole matrix elements, boosting the nonlinear coefficients\cite{meng_dissipative_2021, houver_giant_2019, friedli_four-wave_2013} and allows for tuning the linewidth enhancement factor. Intersubband-based devices have short upper laser state lifetimes, which prevents passive mode-locking for pulse formation. It is therefore essential to model these structures taking into account their fast gain dynamics, in order to find parameter regimes where pulses can be spontaneously formed in ring cavities.

Further investigation will reveal the limits of the widths of these pulses, since they also depend on the size of the ring cavity. Quantitative results require a more intricate treatment of the frequency-dependence of the gain, dispersion, and nonlinearities, which can be accomplished by a formulation in the frequency domain\cite{meng_dissipative_2021}. However, such simulations suffer from difficulties in efficiently exploring their large parameter space. Being able to approximately map such complex models onto the CGLE as described in Appendix \ref{app:realistic} will serve as an important starting point for exploring more general and realistic devices.

\section{Acknowledgements}
Helpful discussions with Bo Meng, Sergej Markmann, Mathieu Bertrand, and J\'er\^ome Faist are gratefully acknowledged. Financial support from the Qombs project funded by the European Union’s Horizon 2020 research and innovation programme under grant agreement no. 820419 is acknowledged.  The simulations were carried out on the Euler computer cluster of ETH Zürich.

\begin{appendix}

\section{Derivation of the CGLE} \label{app:CGLE_derivation}

Our starting point for deriving the equations of motion for the electric field $E$ in our nonlinear system, is the polarization
\begin{eqnarray}
    P(t) &=& \varepsilon_0\int_{-\infty}^t \chi^{(1)}(t, t_1)E(t_1) dt_1
    \\&+& \varepsilon_0\int_{-\infty}^t \chi^{(3)}(t, t_1, t_2, t_3)E(t_1)E(t_2)E(t_3) d^3 t_i, \nonumber
\end{eqnarray}
where the spatial dependence has been suppressed for brevity, but follows the same general structure as the time dependence and will give rise to a phase matching condition that can be applied at the end of the calculation. The susceptibilities $\chi^{(N)}$ include the complete retarded medium response to the electric field up to third order, and describe linear gain, gain saturation, and four-wave mixing (we have ignored $\chi^{(2)}$ since its contribution in the frequency range of interest can be neglected) including all medium memory effects.

The equation of motion for the electric field is the wave equation
\begin{equation}
    \frac{\partial^2 E}{\partial t^2} =
    -\frac{1}{\varepsilon_0}\frac{\partial^2 P}{\partial t^2} + c^2\nabla^2E, \label{eq:Maxwell}
\end{equation}
derived from Maxwell's macroscopic equations.
In the slowly varying envelop and Markov approximations, which are used to derive the LLE, this leads to the modified Lugiato-Lefever equation \begin{equation}
    \frac{\partial E}{\partial t} = \frac{1}{2}(g_0 - \alpha_0)E - \text{i} \sigma_0 \chi^{(3)}_0|E|^2E + \text{i}\frac{c^2}{\omega_0n_0^2}\nabla^2 E
    \label{eq:app_CGLE}
\end{equation}
with gain $g_0$, loss $\alpha_0$, complex nonlinear coefficient $\sigma_0\chi^{(3)}_0$, where $\sigma_0 = 3c\hbar\omega_0^2/(2n_0^3V_0)$. Here, subscripts with 0 indicate quantities evaluated at the central frequency of the spectrum. In this Markovian approach, which avoids convoluted time-integrals and keeping track of fields and polarizations at previous times, the memory effects are lost and so are the dynamics of gain saturation. However, we can easily recover a phenomenological parabolic gain profile by supplementing \eqref{eq:app_CGLE} with a spectral filtering term as in Eqs.~\eqref{eq:general_CGLE} and \eqref{eq:spect_filt}.

\section{Exact soliton solutions}

In the spirit of Ref.~\cite{pereira_nonlinear_1977} we employ the Ansatz of the form
\begin{equation}
    \mathcal{E} = L \text{sech}(K\Theta)^ce^{-i\Omega t}.
\end{equation}
Inserting into Eq.~\eqref{eq:norm_CGLE}
\begin{eqnarray}
0 &=& (1+i\Omega) \mathcal{E} - (1+i\alpha)|L|^2\text{sech}(K\Theta)^{c + c^\ast} \mathcal{E} \\
&+& (1 + i\beta)K^2c\left(c -(c+1)\text{sech}(K\Theta)^2 \right)\mathcal{E}.
\end{eqnarray}
Collecting the exponents of sech, $2\Delta = 2$, and we define $c \equiv 1 + ia$, where $a$ is a real constant. Furthermore, the prefactors to $\text{sech}^2 \mathcal{E}$ and $\mathcal{E}$ yield
\begin{eqnarray}
    (1 + i\alpha) |L|^2 + K^2(1+ia)(2+ia)(1 + i\beta) &=& 0 \\
    1 + i\Omega + K^2(1 + ia)^2(1 + i\beta) &=& 0.
\end{eqnarray}
Separating into real and imaginary parts, we obtain the system of equations
\begin{eqnarray}
\frac{|L|^2}{K^2} + 2 - a^2 - 3\beta a &=& 0 \\
\alpha\frac{|L|^2}{K^2} - a^2\beta + 2\beta + 3a &=& 0 \\
1 + K^2(2 - a^2 - 3\beta a) &=& 0 \\
\Omega + K^2(2a + \beta - a^2\beta) &=& 0.
\end{eqnarray}
The first two equations give
\begin{eqnarray}
a &=& - B \pm \sqrt{B^2 + 2} \\
B &=& \frac{3}{2}\frac{\alpha\beta + 1}{\alpha - \beta},
\end{eqnarray}
whereas the third and fourth provide
\begin{eqnarray}
    K^2 = (a^2 + 2\beta a -1)^{-1} \\
    \Omega = \frac{a^2\beta - \beta - 2a}{(a^2 - 2\beta a -1)}.
\end{eqnarray}
Finally,
\begin{equation}
    |L|^2 = K^2(a^2 + 3\beta a -2) = \frac{(a^2 + 3\beta a -2)}{(a^2 + 2\beta a - 1)}.
\end{equation}
The conditions for this Ansatz giving physical results are $K^2 > 0$ and $|L|^2 > 0$, and there are two choices for $a$ to be checked for each combination of $(\alpha, \beta)$.

\section{Relation to realistic devices}\label{app:realistic}

In order to connect the presented results in terms of normalized parameters to real device properties, the parameters $\beta$ and $\alpha$ are here derived for a general gain medium with a Lorentzian line shape
\begin{equation}
    g(\omega) = g_0\frac{\gamma^2}{(\omega - \omega_0)^2 + \gamma^2},
\end{equation}
where $g_0$ is the unsaturated gain rate, dispersion
\begin{equation}
    \omega(k) \approx \omega_0 + \frac{\partial \omega}{\partial k} k + \frac{1}{2}\frac{\partial^2 \omega}{\partial k^2} k^2,
\end{equation}
and gain saturation
\begin{equation}
    g(I) = \frac{g(0)}{1 + I/I_\text{sat}}.
\end{equation}
By Taylor expansion, we then obtain
\begin{eqnarray}
\beta &\approx& -\frac{c}{2n}\frac{\gamma^2 \text{GVD}}{g_s} \\
\alpha &\approx& -\text{LEF}
\end{eqnarray}
where $g_s = g_\text{th}(2 - g_\text{th}/g_0)$, $g_\text{th}$ is the threshold gain, and LEF is the linewidth enhancement factor. This is also roughly equivalent to a Maxwell-Bloch model under certain conditions.\cite{piccardo_frequency_2020}

For a typical mid-infrared QCL, LEF$\approx 0.5-1$, GVD$\approx - 2000$ fs$^2$/mm, $g_s \approx 20$ ns$^{-1}$, $\gamma \approx 10$ ps$^{-1}$, $n \approx 3.2$. This gives $\beta \approx 0.7$. From Fig.~\ref{fig:bistability_2d}c) it is clear that these parameters cannot give rise to amplitude modulated solutions. However, by design it is possible to lower slightly the GVD and the threshold gain, and increase $\gamma$ (by e.~g.~making a broad gain including several transitions), so that large values of $\beta$ are also reachable. For instance, $\gamma = 15$ ps$^{-1}$, $g_s = 10$ ns$^{-1}$, and GVD $= -7000$ fs$^2$/mm gives $\beta \approx 7$, which would allow the region (1) in Fig.~\ref{fig:analytical_width}a) to be reached. Also large LEF can be achieved in QCLs. As an example, taking the values of Ref.~\cite{piccardo_frequency_2020}, we get $\alpha = -2$ and $\beta = 0.5$ which is precisely at the edge of the BF instability, in accordance with the full Maxwell-Bloch theory\cite{piccardo_frequency_2020}. Our simplified simulations correctly reproduces the low amplitude modulation observed in that experiment, as seen by solution number (4) in Fig.~\ref{fig:analytical_width} a). On the other hand, the simulations in Ref.~\cite{meng_dissipative_2021} suggest $\text{LEF}\approx4$, and allows for significantly amplitude-modulated solitons, which were however difficult to reproduce (and hence not shown in Fig.~\ref{fig:analytical_width}).

Also negative LEF can be realized by designing the active quantum well structure or the waveguide. In this case, a positive GVD can be used, which is more easily achievable than a negative one.

In the case of a THz QCL, operating below the LO phonon frequency and with higher waveguide losses means that $\gamma$ is lower than for mid-IR QCLs, while $g_s$ is higher. On the other hand, the laser operates much closer to the TO phonon resonance and is more sensitive to the cavity geometry, which can yield very large GVD values up to $\pm 10^5$ fs$^2$/mm\cite{bachmann_dispersion_2016}. This high GVD could compensate and allow soliton operation also in THz ring QCLs\cite{jaidl_comb_2021}.

Interband lasers have larger LEF than intersubband ones in general ($\text{LEF}\sim3.5$)\cite{sorel_operating_2003}, and frequency-modulated combs have been achieved recently in monolithic devices\cite{sterczewski_frequency-modulated_2020}. However, so far monolithic ring interband lasers have only demonstrated single-mode operation\cite{sorel_operating_2003, born_lasing_2008}.
Quantum dot ring lasers have also demonstrated multi-mode operation with very low threshold current\cite{zhang_hybrid_2019}, and with similar LEF of mid-infrared QCLs\cite{huang_analysis_2018} are also promising candidates for gain solitons provided large enough dispersion can be achieved. It should be kept in mind that, strictly speaking, the derived model is only valid for instantaneous gain saturation, and therefore its applicability to inter-band lasers has yet to be investigated.

\end{appendix}

\end{document}